\documentclass[onecolumn,showpacs]{revtex4}
\usepackage{amssymb}
\usepackage{makeidx}
\usepackage{amsmath}
\usepackage{graphicx}
\usepackage{epstopdf}
\usepackage{dcolumn}
\usepackage{bm}

\input{tcilatex}
\begin{document}

\title{Third harmonic generation in gapped bilayer graphene }
\author{A.K. Avetissian}
\author{A.G. Ghazaryan}
\author{Kh. V. Sedrakian}
\affiliation{Centre of Strong Fields, Yerevan State University, 1 A. Manukian,
Yerevan 0025, Armenia }
\date{\today }

\begin{abstract}
With the help of numerical simulations in microscopic nonlinear quantum
theory of coherent electromagnetic radiation interaction with a gapped
bilayer graphene, we find out the optimal values of pump wave intensity,
graphene temperature, and energy gap induced by a constant electric field
for practically significant third order harmonic coherent emission. The
Liouville-von Neumann equation is treated numerically for the third harmonic
generation in multiphoton excitation regime near the Dirac points of the
Brillouin zone. We examine the rates of the third harmonic at the
particle-hole annihilation in the field of a strong pump wave of linear
polarization for practically real/optimal parameters of a considering
system. The obtained results show that by choosing the optimal values of the
main characteristic parameters, a gapped bilayer graphene can serve as an
effective medium for generation of the third harmonic at room temperatures
in the terahertz and far infrared domains.
\end{abstract}

\maketitle



\section{Introduction}

Recently the interest has grown to harmonics generation and related
multiphoton processes in graphene and other 2D nanostructures with the
similar properties \cite%
{1,2,4,5,7,24,25,26,27,28,29,30,31,32,33,35,36,37,38,38a,38b} due to the
achievements of production of new nanomaterials. Many experiments already
exist where the theoretical predictions have been justified, in particular,
the experiment \cite{36} with generation of $9$th harmonic in graphene in
infrared regime, and generation of harmonics extending to the $13$th order
in 2D semiconductor \cite{41} and in solids \cite{42}. In these
nanostructures at very high carrier mobility, with increasing pump wave
intensity one can achieve an ultrafast excitation regime and it is important
that threshold values of intensities of electromagnetic (EM) radiation for
observation of nonlinear phenomena compared with free electrons and atoms
are essentially reduced. In particular, for the infrared photons one can
obtain nonlinear effects at the $10^{9}$ times smaller intensities of an
external EM field \cite{Book}.

Bilayer graphene ($AB$ stacked) \cite{2,1a,1aa,27} (as the other multilayer
2D nanostructures \cite{1,2,4,5,7}) is of great interest since its
electronic states are considerably richer than those of a monolayer graphene 
\cite{9,22b,24b,25b}. But both single- and multilayer graphene in their
unperturbed state lack the band gap \cite{19,20}. However, multilayer
graphene materials under the application of a perpendicular electric field
may exhibit the ability to produce an electrically tunable band gap. These
outcomes disclose new possibilities of a much wider range of applications
for graphene materials in electronics and photonics \cite{9,10,14,16,8a}.
Moreover, in the case of$\ AB$ stacked bilayer graphene the trigonal warping
effect for the energy spectrum deforms the low-energy{\LARGE \ }excitation
with $\mathcal{E}\lesssim 10$ $\mathrm{meV}$, which significantly enhances
the rates of the harmonics \cite{27} in the terahertz region compared to a
monolayer graphene. It takes place the modification of quasi-energy
spectrum, in particular, the induction of valley polarized currents \cite%
{26b,27b}, as well as second- and third- order nonlinear-optical effects 
\cite{28b,29b,30b,Exp,31b} in $AB$ stacked bilayer graphene under the
applied intense EM field. So, bilayer graphene ($AB$ stacked) has some
advantage as compared to a single layer graphene for the photonic
applications because of its anisotropic zone structure and the extensive
tunable band gap \cite{1a,1aa,3a,4a,5a,6a,7a,10a,11a}. The large tunable
band gaps with maximal energy $U\simeq 280$ $\mathrm{meV}$ \cite{16} can
make possible to control nonlinear EM response in bilayer graphene at room
temperatures, which is impossible in intrinsic graphene \cite{27}.

The aim of the current paper is to find out optimal parameters of\
light-gapped bilayer graphene interaction for third order harmonic
radiation. We investigate third order harmonic radiation efficiency
depending on the value of energy gap, temperature, pump wave frequency, and
intensity. The consideration is based on the solution of the Liouville-von
Neumann equation for density matrix in multiphoton excitation regime near
the Dirac points of the Brillouin zone. We show that there is an extensive
emission (coherent) of the third harmonic at the wave-induced particle or
hole acceleration and annihilation. We have revealed the optimal conditions
for third harmonic generation.

The paper organization is the following. In Sec. II the equations for a
single-particle density matrix is numerically solved in the multiphoton
interaction regime. In Sec. III, the problem of third order harmonic
generation at the multiphoton excitation of gapped bilayer graphene is
considered. The conclusions are given in Sec. IV.

\section{Basic model and theory}

Back to the issue of the third harmonic radiation let us choose initially
the interaction parameters as follow. At an intraband transitions the
interaction of a particle with the wave at the photon energies $\hbar \omega
>\mathcal{E}_{L}$\ \cite{27} characterizes by the known dimensionless
parameter $\chi $:%
\begin{equation}
\chi =eE_{0}/(\omega \sqrt{m\hbar \omega }),  \label{00}
\end{equation}%
where $E_{0}$ is a wavestrength, $\omega $ is a wavefrequency, $e$ is an
electron charge and $m$ is an effective mass. Moreover, due to the gap the
interband transitions are characterized by so-called Keldysh \cite{Keld}
parameter 
\begin{equation}
\gamma =\omega \sqrt{mU}/\left( eE_{0}\right) =\chi ^{-1}\sqrt{U/\left(
\hbar \omega \right) }.  \label{01}
\end{equation}%
Here $U$ is a band gap energy, $\mathcal{E}_{L}$ is the Lifshitz energy, $%
\hbar $ is the Planck constant. The last defines the tunneling ($\gamma <<1$%
) or multiphoton ($\gamma >>1$) character of the ionization process in the
strong laser field. For the considered case, the ionization process reduces
to the transfer of the electron from the valence band into the conduction
band that is the creation of an electron-hole pair. Since the interband
transitions can be neglected when $\gamma >>1$, then the wave field cannot
provide enough energy for the creation of an electron-hole pair, and the
generation of harmonics is suppressed. If $\gamma \sim 1$ or $\gamma <<1$,
interband transitions take place. In the current paper, we will consider the
nonadiabatic regime for the generation of third harmonic at $\chi \sim 1$
and $\gamma \sim 1$.

\begin{figure}[tbp]
\includegraphics[width=.51\textwidth]{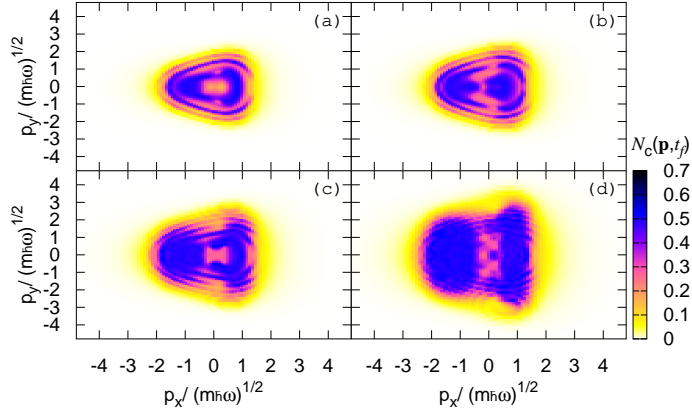}
\caption{(Color online) Creation of a particle-hole pair in bilayer graphene
at multiphoton resonant excitation with the trigonal warping effect for
middle gap energy $U=0.1$\textrm{\ }$\mathrm{eV}$. Particle distribution
function $N_{c}(\mathbf{p},t_{f})$\ (in arbitrary units) after the
interaction is displayed at various wave frequencies: (a) $\protect\omega =$%
\ $0.06$\ $\mathrm{eV/}\hbar $, (b) $\protect\omega =$\ $0.05$\ $\mathrm{eV/}%
\hbar $, (c) $\protect\omega =$\ $0.04$\ $\mathrm{eV/}\hbar $, and (d) $%
\protect\omega =$\ $0.03$\ $\mathrm{eV/}\hbar $. The temperature is taken to
be $T=0.025$ $\mathrm{eV}$, the wave intensity is $I_{\protect\chi %
}=7.5\times 10^{6}\mathrm{\ Wcm}^{-2}$. The results are for the valley $%
\protect\zeta =1$. }
\end{figure}

Let us investigate the multiphoton excitations of Fermi-Dirac sea in the
gapped bilayer graphene in nonadiabatic regime. In accordance with the
microscopic nonlinear quantum theory of $AB$ stacked gapped bilayer
graphene, the low-energy excitations $\left\vert \mathcal{E}_{\sigma
}\right\vert <\gamma _{1}\simeq 0.39$ $\mathrm{eV}$ in the immediate
vicinity of the Dirac points $K_{\zeta }$ (valley quantum number $\zeta =\pm
1$) one can describe by the effective single particle Hamiltonian \cite%
{9,22b,24b}:

\begin{equation}
\widehat{H}_{\zeta }=\left( 
\begin{array}{cc}
\frac{U}{2} & h_{\zeta }^{\ast }\left( \mathbf{p}\right) \\ 
h_{\zeta }\left( \mathbf{p}\right) & -\frac{U}{2}%
\end{array}%
\right) .  \label{1}
\end{equation}%
Here $\mathbf{\hat{p}}=\left\{ \widehat{p}_{x},\widehat{p}_{y}\right\} $%
\textbf{\ }is the electron momentum operator, $m=\gamma _{1}/(2\mathrm{v}%
_{F}^{2})$ is the effective mass ($\mathrm{v}_{F}$ is the Fermi velocity in
a monolayer graphene); $\mathrm{v}_{3}=\sqrt{3}b\gamma _{3}/(2\hbar )$ is
the effective velocity related to oblique interlayer hopping $\gamma
_{3}=0.32$ $\mathrm{eV}$ ($b\approx 0.246$ $\mathrm{nm}$ is the distance
between the nearest $A$ sites); and 
\begin{equation}
h_{\zeta }\left( \mathbf{p}\right) =-\frac{1}{2m}\left( \zeta \widehat{p}%
_{x}+i\widehat{p}_{y}\right) ^{2}+\mathrm{v}_{3}\left( \zeta \widehat{p}%
_{x}-i\widehat{p}_{y}\right) .  \label{2}
\end{equation}%
The diagonal elements in Eq. (\ref{1}) correspond to induced by
perpendicular to graphene plane electric field gap $U$. The first term in
Eq. (\ref{2}) gives a pair of parabolic bands $E=\pm p^{2}/(2m)$, and the
second term connects with $\gamma _{3}$ causes trigonal warping in the band
dispersion. The Lifshitz transition (separation of the Fermi surface) in the
low-energy region occurs at an energy $\mathcal{E}_{L}=m\mathrm{v}%
_{3}^{2}/2\simeq 1$ $\mathrm{meV}$, and the two touching parabolas are
transformed into the four separate \textquotedblleft
pockets\textquotedblright\ \cite{27}. The spin and the valley quantum
numbers are conserved. There is no degeneracy upon the valley quantum number 
$\zeta $. Since there are no intervalley transitions, the valley index $%
\zeta $ one has considered as a parameter.

\begin{figure}[tbp]
\includegraphics[width=.51\textwidth]{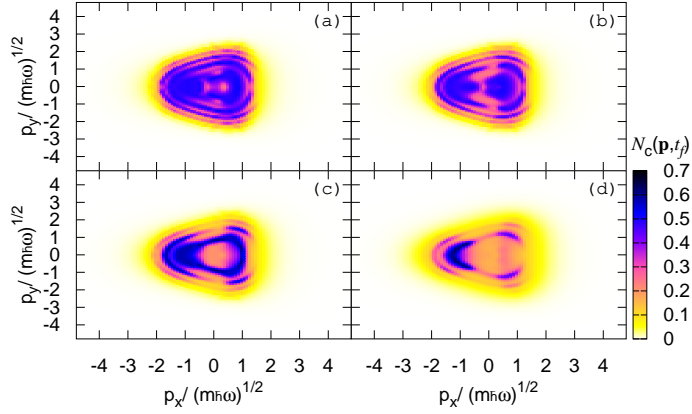}
\caption{ (Color online) The same as Fig. 1 but for fixed dimensionless
intensity parameter $\protect\chi =1$. The results are for the valley $%
\protect\zeta =1$: (a)--(d) correspond to the photon energy $\hbar \protect%
\omega \mathrm{/eV}=$\ $0.06$, $0.05$, $0.04$ and $0.03$, respectively. }
\end{figure}

The eigenstate functions of the effective Hamiltonian (\ref{1}) are the
spinor ones,%
\begin{equation}
\Psi _{\sigma }(\mathbf{r})=\frac{1}{\sqrt{S}}|\sigma ,\mathbf{p}\rangle e^{%
\frac{i}{\hbar }\mathbf{pr}}  \label{3}
\end{equation}%
with%
\begin{equation}
|\sigma ,\mathbf{p}\rangle =\frac{1}{\sqrt{S}}\sqrt{\frac{\mathcal{E}%
_{\sigma }+\frac{U}{2}}{2\mathcal{E}_{\sigma }}}\left( 
\begin{array}{c}
1 \\ 
\frac{1}{\mathcal{E}_{\sigma }+\frac{U}{2}}\Gamma \left( \mathbf{p}\right)%
\end{array}%
\right) .  \label{4}
\end{equation}%
Here 
\begin{equation}
\mathcal{E}_{\sigma }\left( \mathbf{p}\right) =\sigma \sqrt{\frac{U^{2}}{4}%
+\left( \mathrm{v}_{3}p\right) ^{2}-\zeta \frac{\mathrm{v}_{3}p^{3}}{m}\cos
3\vartheta +\left( \frac{p^{2}}{2m}\right) ^{2}};  \label{5}
\end{equation}%
are the corresponding eigenenergies, and 
\begin{equation}
\Gamma \left( \mathbf{p}\right) =-\frac{p^{2}}{2m}e^{i2\zeta \vartheta
}+\zeta \mathrm{v}_{3}pe^{-i\zeta \vartheta },  \label{6}
\end{equation}%
$\vartheta =\arctan \left( p_{y}/p_{x}\right) $, $S$ is the quantization
area, and $\sigma $ is the band index: $\sigma =1$ and $\sigma =-1$ for
conduction and valence bands.

\begin{figure}[tbp]
\includegraphics[width=.45\textwidth]{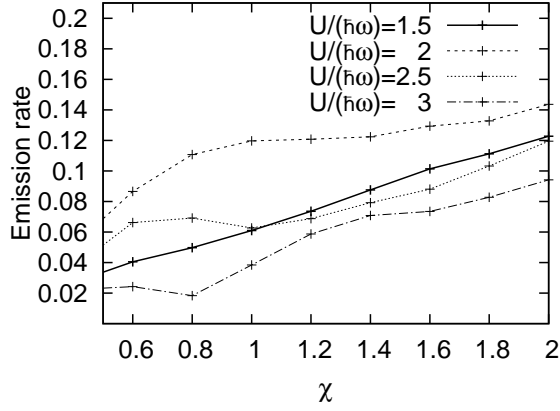}
\caption{ Third harmonic emission rate $J_{3}$ (in arbitrary units) for
gapped bilayer graphene versus $\protect\chi $ for various band gaps. The
temperature is taken to be $T=0.025$\textrm{\ }$\mathrm{eV}$, wave frequency
is $\protect\omega =$\ $0.05$\ $\mathrm{eV/}\hbar $. }
\end{figure}

To exclude the effect of the magnetic field, the EM wave propagates in the
perpendicular direction to the gapped bilayer graphene sheets ($XY$) . A
plane quasimonochromatic EM wave is taken to be linearly polarized along the 
$Y$ axis: 
\begin{equation}
\mathbf{E}\left( t\right) =f\left( t\right) E_{0}\widehat{\mathbf{y}}\sin
\omega t.  \label{7}
\end{equation}%
The wave slowly varying envelope $f\left( t\right) $ is defined by the
function:%
\begin{equation}
f\left( t\right) =\left\{ 
\begin{array}{cc}
\sin ^{2}\left( \pi t/\mathcal{T}_{p}\right) , & 0\leq t\leq \mathcal{T}_{p},
\\ 
0, & t<0,t>\mathcal{T}_{p},%
\end{array}%
\right. ,  \label{8}
\end{equation}%
where $\mathcal{T}_{p}=20\mathcal{T}$ is the pulse duration, and $\mathcal{T}%
=2\pi /\omega $.

By the second quantization formalism, the fermionic field operator,
expanding on the basis of free states (\ref{3}), can be written in the form: 
\begin{equation}
\widehat{\Psi }(\mathbf{r},t)=\sum\limits_{\mathbf{p,}\sigma }\widehat{a}_{%
\mathbf{p},\sigma }(t)\Psi _{\sigma }(\mathbf{r}).  \label{9}
\end{equation}%
Here $\widehat{a}_{\mathbf{p},\sigma }(t)$ and $\widehat{a}_{\mathbf{p}%
,\sigma }^{+}(t)$ are the annihilation and creation operators for an
electron with the momentum $\mathbf{p}$, which satisfy the fermionic
anticommutation rules at equal times. The single-particle Hamiltonian in the
presence of a uniform time-dependent electric field $E(t)$ in a length gauge
can be expressed in the form:%
\begin{equation}
\widehat{H}_{s}=\widehat{H}_{\zeta }+\left( 
\begin{array}{cc}
e\mathbf{rE}\left( t\right) & 0 \\ 
0 & e\mathbf{rE}\left( t\right)%
\end{array}%
\right) .  \label{12}
\end{equation}%
Using the expansion (\ref{9}), the second quantized total Hamiltonian can be
presented in the form: 
\begin{equation}
\widehat{H}=\sum\limits_{\sigma ,\mathbf{p}}\mathcal{E}_{\sigma }\left( 
\mathbf{p}\right) \widehat{a}_{\sigma \mathbf{p}}^{+}\widehat{a}_{\sigma 
\mathbf{p}}+\widehat{H}_{\mathrm{int}},  \label{13}
\end{equation}%
where the part of interaction with the EM\ field is given by the relation:%
\begin{equation*}
\widehat{H}_{\mathrm{int}}=ie\sum\limits_{\mathbf{p,p}^{\prime },\sigma
}\delta _{\mathbf{p}^{\prime }\mathbf{p}}\partial _{\mathbf{p}^{\prime }}%
\mathbf{E}\left( t\right) \widehat{a}_{\mathbf{p},\sigma }^{\dagger }%
\widehat{a}_{\mathbf{p}^{\prime },\sigma ^{\prime }}
\end{equation*}%
\begin{equation}
+\sum\limits_{\mathbf{p},\sigma }\mathbf{E}\left( t\right) \left( \mathbf{D}%
_{\mathrm{t}}\left( \sigma ,\mathbf{p}\right) \widehat{a}_{\mathbf{p},\sigma
}^{+}\widehat{a}_{\mathbf{p},-\sigma }+\mathbf{D}_{\mathrm{m}}\left( \sigma ,%
\mathbf{p}\right) \widehat{a}_{\mathbf{p},\sigma }^{+}\widehat{a}_{\mathbf{p}%
,\sigma }\right) ,  \label{14}
\end{equation}%
with the transition dipole moment 
\begin{equation}
\mathbf{D}_{\mathrm{t}}\left( \sigma ,\mathbf{p}\right) =\hbar e\langle
\sigma ,\mathbf{p}|i\partial _{\mathbf{p}}|-\sigma ,\mathbf{p}\rangle
\label{15}
\end{equation}%
and the mean dipole moment or Berry connection: 
\begin{equation}
\mathbf{D}_{\mathrm{m}}\left( \sigma ,\mathbf{p}\right) =\hbar e\langle
\sigma ,\mathbf{p}|i\partial _{\mathbf{p}}|\sigma ,\mathbf{p}\rangle .
\label{16}
\end{equation}

\begin{figure}[tbp]
\includegraphics[width=.43\textwidth]{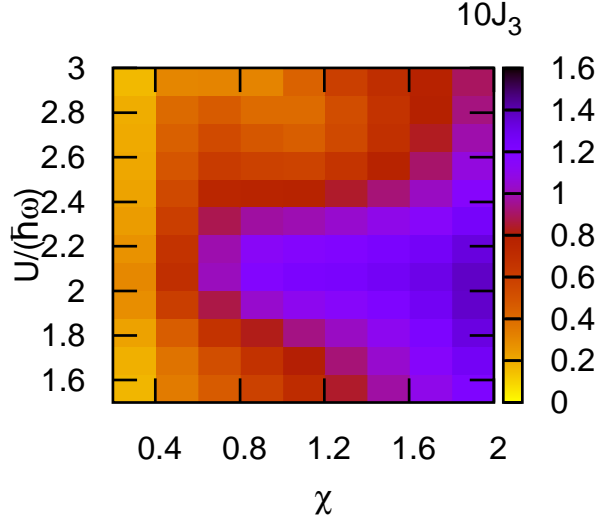}
\caption{ $3$rd harmonic emission rate in bilayer graphene at multiphoton
excitation $J_{3}$ (in arbitrary units) versus the parameter $\protect\chi $
and the gap energy $U/\left( \hbar \protect\omega \right) $ at the same
other parameters as in Fig. 3. }
\end{figure}

The ultimate expressions of the dipole moment components are given in the
Appendices by Eqs. (\ref{27})--(\ref{30}).

For description of multiphoton interaction in a gapped bilayer graphene the
Liouville--von Neumann equation with a single-particle density matrix 
\begin{equation}
\rho _{\alpha ,\beta }(\mathbf{p},t)=\langle \widehat{a}_{\mathbf{p},\beta
}^{+}\left( t\right) \widehat{a}_{\mathbf{p},\alpha }\left( t\right) \rangle
\label{17}
\end{equation}
has the form: 
\begin{equation}
i\hbar \frac{\partial \widehat{a}_{\mathbf{p},\alpha }\left( t\right) }{%
\partial t}=\left[ \widehat{a}_{\mathbf{p},\alpha }\left( t\right) ,\widehat{%
H}\right] ,  \label{18}
\end{equation}%
where $\widehat{a}_{\mathbf{p},\alpha }\left( t\right) $ obeys the
Heisenberg equation.

We use only the $\mathbf{p}$-diagonal elements of the density matrix due to
the homogeneity of the problem. Since homogeneous relaxation processes are
slow compared with inhomogeneous ones, the relaxation processes can be taken
into Liouville--von Neumann equation including the inhomogeneous
phenomenological damping term. Hence, using the Eqs. (\ref{13})-(\ref{18}),
we have to solve the differential equations 
\begin{equation*}
i\hbar \frac{\partial \rho _{\alpha ,\beta }(\mathbf{p},t)}{\partial t}%
-i\hbar e\mathbf{E}\left( t\right) \frac{\partial \rho _{\alpha ,\beta }(%
\mathbf{p},t)}{\partial \mathbf{p}}=
\end{equation*}%
\begin{equation*}
\left( \mathcal{E}_{\alpha }\left( \mathbf{p}\right) -\mathcal{E}_{\beta
}\left( \mathbf{p}\right) -i\hbar \Gamma \left( 1-\delta _{\alpha \beta
}\right) \right) \rho _{\alpha ,\beta }(\mathbf{p},t)
\end{equation*}%
\begin{equation*}
+\mathbf{E}\left( t\right) \left( \mathbf{D}_{\mathrm{m}}\left( \alpha ,%
\mathbf{p}\right) -\mathbf{D}_{\mathrm{m}}\left( \beta ,\mathbf{p}\right)
\right) \rho _{\alpha ,\beta }(\mathbf{p},t)
\end{equation*}%
\begin{equation}
+\mathbf{E}\left( t\right) \left[ \mathbf{D}_{\mathrm{t}}\left( \alpha ,%
\mathbf{p}\right) \rho _{-\alpha ,\beta }(\mathbf{p},t)-\mathbf{D}_{\mathrm{t%
}}\left( -\beta ,\mathbf{p}\right) \rho _{\alpha ,-\beta }(\mathbf{p},t)%
\right] ,  \label{19}
\end{equation}%
where $\Gamma $ is the damping rate. The particle distribution functions for
the conduction $N_{c}(\mathbf{p},t)=\rho _{1,1}(\mathbf{p},t)$ and valence $%
N_{\mathrm{v}}(\mathbf{p},t)=\rho _{-1,-1}(\mathbf{p},t)$ bands are
correspond to diagonal elements. The nondiagonal elements are interband
polarization $\rho _{1,-1}(\mathbf{p},t)=P(\mathbf{p},t)$ and its complex
conjugate $\rho _{-1,1}(\mathbf{p},t)=P^{\ast }(\mathbf{p},t)$.

As an initial state, we present an ideal Fermi gas in equilibrium state with
the chemical potential to be zero. We need to solve the set of Eqs. (\ref{19}%
), and followed from the last closed set of differential equations (\ref{20}%
)--(\ref{21a}) given in the Appendices for the quantities $N_{\mathrm{v}}(%
\mathbf{p},t)$, $N_{c}(\mathbf{p},t)$, $P(\mathbf{p},t)$, taking into
account the initial conditions:%
\begin{equation}
P(\mathbf{p},0)=0;N_{c}(\mathbf{p},0)=\frac{1}{1+e^{\mathcal{E}_{1}\left( 
\mathbf{p}\right) /T}},  \label{22}
\end{equation}%
\begin{equation}
N_{\mathrm{v}}(\mathbf{p},0)=1-N_{c}(\mathbf{p},0),  \label{23}
\end{equation}%
where $T$ is the temperature in energy units.

\begin{figure}[tbp]
\includegraphics[width=.45\textwidth]{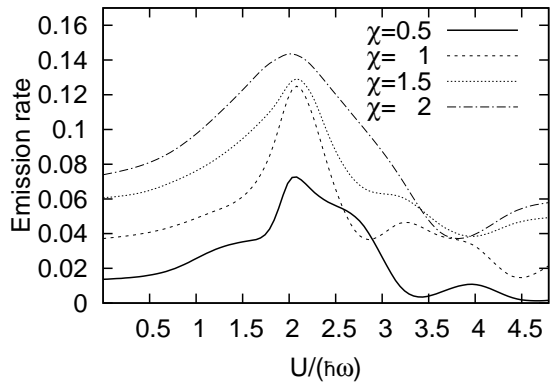}
\caption{ $3$rd harmonic emission rate in bilayer graphene $J_{3}$ (in
arbitrary units), as a function of the gap energy scaled to photon energy
for various wave intensities. The temperature is taken to be $T=0.025$%
\textrm{\ }$\mathrm{eV}$. The wave frequency is $\protect\omega =$\ $0.05$\ $%
\mathrm{eV/}\hbar $. }
\end{figure}

For the numerical solution of the set of equations (\ref{20})--(\ref{21a})
we transformed to the equations with partial derivatives into the ordinary
ones. We changed the variables by new ones $t$ and $\widetilde{\mathbf{p}}=%
\mathbf{p}-\mathbf{p}_{E}(t)$, where $\mathbf{p}_{E}(t)=-e\int_{0}^{t}%
\mathbf{E}\left( t^{\prime }\right) dt$. The last is the classical momentum
given by the wave field, which with the Keldish parameter (\ref{01}) and
wave intensity characterize the intraband transitions. The integration of
equations (\ref{20})--(\ref{21a}) is performed on a homogeneous grid of $%
10^{4}$ ($\widetilde{p}_{x},\widetilde{p}_{y}$)-points. For the maximal
momentum we take $\widetilde{p}_{\max }/\sqrt{m\hbar \omega }=5$. The time
integration is performed by the fourth-order adaptive Runge-Kutta method.
The relaxation rate is given $\Gamma =0.5\mathcal{T}^{-1}$.

For all calculations an EM wave is taken to be linearly polarized along the $%
Y$\ axis with frequencies in terahertz domain ($1.24-124$ $\mathrm{meV}$).
Similar calculations for a wave linearly polarized along the $X$ axis show
qualitatively the same picture. Note that the intensity of the wave can be
estimated as 
\begin{equation}
I_{\chi }=\chi ^{2}\times 6\times 10^{10}\mathrm{Wcm}^{-2}(\hbar \omega /%
\mathrm{eV})^{3},  \label{23a}
\end{equation}%
so the the multiphoton effects at $\chi \simeq 1$ for terahertz frequencies
become essential at $I_{\chi }=10^{2}-10^{8}\mathrm{\ Wcm}^{-2}$. If in case
of $\gamma <<1$ the process of tunneling transitions is independent on the
wave frequency, in considered cases the value $I_{\chi }$ required for the
nonlinear regime strongly depends on the photon energy. The case of
multiphoton transitions, depending on the wave frequency is considered
separately.

In Figs. 1, 2, the photoexcitations of the Fermi-Dirac sea are presented at
room temperatures. In Fig. 1, the density plot of the particle distribution
function $N_{c}(\mathbf{p},t_{f})$\ is shown as a function of the scaled
dimensionless momentum components after the interaction at the same wave
intensity $I_{\chi }$ for various frequencies. The pulse duration is $%
\mathcal{T}_{p}=20\mathcal{T}\approx 1.6$ $\mathrm{ps}$, and the intensity
is taken to be same: $I_{\chi }\simeq 10^{7}\mathrm{\ Wcm}^{-2}$. We have
taken the middle band gap $U\simeq 0.1$ $\mathrm{eV}$, because as seen from
similar calculations for a large band gap, with the increase of $U$ we
approach to perturbation regime at $\gamma >1$, and only a weak excitation
of Fermi-Dirac sea is appeared. In Fig. 2, for the various pump wave
frequencies the photoexcitation dependence on the pump wave intensity is
presented at the same dimensionless intensity parameter $\chi =1$, for the
same other parameters ($U,T$) as in Fig. 1. In viewed figures with the
legible trigonal warping effect, describing the deviation of the excited
iso-energy contours from circles, at the values of $\chi \succsim 1$ when $%
\gamma \simeq 1$ the multiphoton excitations have clearly seen. As is known,
the multiphoton excitation of the Fermi-Dirac sea takes place along the
trigonally warped isolines of the quasienergy $\mathcal{E}_{1}\left( 
\widetilde{\mathbf{p}}+\mathbf{p}_{E}\left( t\right) ,t\right) $ spectrum
modified by the wave field. Thus, the multiphoton probabilities of
particle-hole pair production will have maximal values for the iso-energy
contours defined by the resonant conditions: $\mathcal{T}^{-1}\int%
\limits_{0}^{\mathcal{T}}2\mathcal{E}_{1}\left( \widetilde{\mathbf{p}}+%
\mathbf{p}_{E}\left( t\right) ,t\right) dt=n\hbar \omega \ $($n$ is a
natural number). Further, Figs. 1, 2 demonstrate the Fermi-Dirac sea
photoexcitations strong dependence on the pump wave frequency. As is seen
from the last considerations, in Fig. 1 with a fixed pump wave intensity (%
\ref{23a}) the states with absorption of more photons are appearing in the
Fermi-Dirac sea with decreasing of the wave frequency. The last takes place
with increasing of the wave frequency in Fig. 2 when the dimensionless
intensity parameter is fixed.

In the following section we will investigate the nonlinear response of a
bilayer graphene in the process of third order harmonic generation under the
influence of laser field in nonadiabatic regime $\gamma \simeq 1$ with the
frequencies in terahertz domain: $\omega =0.03\div 0.06$ $\mathrm{eV/}\hbar $%
.

\begin{figure}[tbp]
\includegraphics[width=.45\textwidth]{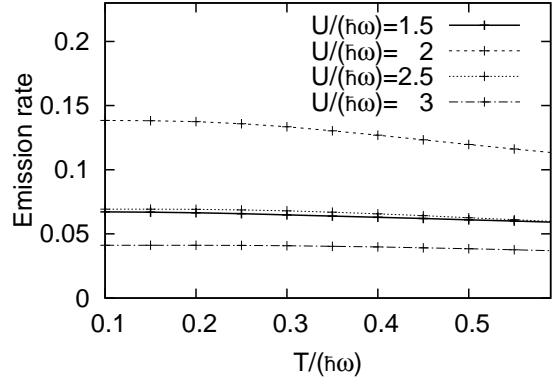}
\caption{ $3$rd harmonic emission rate in bilayer graphene $J_{3}$ (in
arbitrary units) as a function of the temperature scaled to photon energy
for various band gaps. The wave intensity defines by parameter $\protect\chi %
=1$ and frequency $\protect\omega =$\ $0.05$\ $\mathrm{eV/}\hbar $. }
\end{figure}

\section{Numerical simulations for third order harmonic generation in gapped
bilayer graphene}

As was mentioned in Section II, at the multiphoton resonant excitation and
particle-hole annihilation from the coherent superposition states the
intense coherent radiation of the harmonics of the applied EM wave will
occur. Here we consider the possibility of harmonic generation from the
multiphoton excited states depending on the pump field intensity, created
energy gap in a bilayer graphene, and temperature of the initial stationary
state. The coherent part of the radiation spectrum we present by the mean
value of the current density operator in the form:%
\begin{equation}
j_{\zeta }=-eg_{s}\left\langle \widehat{\Psi }(\mathbf{r},t)\left\vert 
\widehat{\mathbf{v}}_{\zeta }\right\vert \widehat{\Psi }(\mathbf{r}%
,t)\right\rangle .  \label{50}
\end{equation}%
Here $g_{s}=2$ is the spin degeneracy factor, $\widehat{\mathbf{v}}_{\zeta }$
is the velocity given by (\ref{51}), (\ref{52}) in the Appendices.

Using the Eqs. (\ref{50}), (\ref{51}), (\ref{52}) and (\ref{17}), the
expectation value of the current for the valley $\zeta $ can be written in
the form:%
\begin{equation*}
\mathbf{j}_{\zeta }\left( t\right) =-\frac{g_{s}e}{(2\pi \hbar )^{2}}\int d%
\mathbf{p}\left\{ \mathbf{V}\left( \mathbf{p}\right) \left( N_{c}(\mathbf{p}%
,t)-N_{\mathrm{v}}(\mathbf{p},t)\right) \right.
\end{equation*}%
\begin{equation}
\left. +2\hbar ^{-1}i\mathcal{E}_{1}\left( \mathbf{p}\right) \left[ \mathbf{D%
}_{\mathrm{t}}\left( \mathbf{p}\right) P^{\ast }(\mathbf{p},t)-\mathbf{D}_{%
\mathrm{t}}^{\ast }\left( \mathbf{p}\right) P(\mathbf{p},t)\right] \right\} ,
\label{53}
\end{equation}%
where 
\begin{equation}
\mathbf{V}\left( \mathbf{p}\right) =\frac{\mathrm{v}_{3}\mathbf{p}-3\zeta 
\frac{\mathrm{v}_{3}p}{2m}\mathbf{p}\cos 3\vartheta +3\zeta \frac{\mathrm{v}%
_{3}p^{3}}{2m}\sin 3\vartheta \frac{\partial \vartheta }{\partial \mathbf{p}}%
+2\frac{\mathbf{p}^{3}}{\left( 2m\right) ^{2}}}{\mathcal{E}_{1}\left( 
\mathbf{p}\right) }  \label{54}
\end{equation}%
is the intraband velocity. The total current is composed from two terms with 
$N_{c}(\mathbf{p},t)$ and $P(\mathbf{p},t)$ which provide two sources for
the generation of harmonics -intraband and interband, respectively.
Interband high harmonics are radiated as a result of pump field-induced
recombination of accelerated electron-hole pairs. Intraband high harmonics
are generated as a result of the independent motion of carriers in their
respective bands. Note that the relative contribution of intraband and
interband high harmonics strongly depends on the nanostructure's material
and pump field parameters. And the contributions of both mechanisms are
essential when the energy gap is small compared with the wave photon energy 
\cite{Avet2019}.

Since there is no degeneracy upon valley quantum number $\zeta $, the total
current is obtained by a summation over $\zeta $, and the last has the form: 
\begin{equation}
j_{x}=j_{1,x}+j_{-1,x};  \label{55}
\end{equation}%
\begin{equation*}
j_{y}=j_{1,y}+j_{-1,y}.
\end{equation*}%
The scaled total current components are the functions:%
\begin{equation}
\frac{j_{x,y}}{j_{0}}=J_{x,y}\left( \omega t,\chi ,\gamma ,\frac{\mathcal{E}%
_{L}}{\hbar \omega },\frac{T}{\hbar \omega },\frac{U}{\hbar \omega }\right) ,
\label{56a}
\end{equation}%
where $j_{0}=\pi ^{-2}e\omega \sqrt{m\omega /\hbar }$, $J_{x}$ and $J_{y}$
are the dimensionless periodic functions in case of monochromatic wave. They
parametrically depend on the interaction parameters $\chi $ (\ref{00}), $%
\gamma $ (\ref{01}), scaled Lifshitz energy, temperature and gap energy.
Thus, having solutions of \ Eqs. (\ref{20})-(\ref{21a}), and making an
integration in Eqs. (\ref{53}), (\ref{54}) one can calculate the $n$th
harmonic radiation spectra with the help of a Fourier transform of the
function $J_{x,y}(t)$. The emission rate of the $n$th harmonic is
proportional to $n^{2}|j_{n}|^{2}$, where 
\begin{equation}
|j_{n}|^{2}=|j_{xn}|^{2}+|j_{yn}|^{2}.  \label{57}
\end{equation}%
Here $j_{xn}$ and $j_{yn}$ are $n$th Fourier components of the field-induced
total current for which the fast Fourier transform algorithm has been used.
For the all plots one has used the normalized current density (\ref{56a}).

\begin{figure}[tbp]
\includegraphics[width=.45\textwidth]{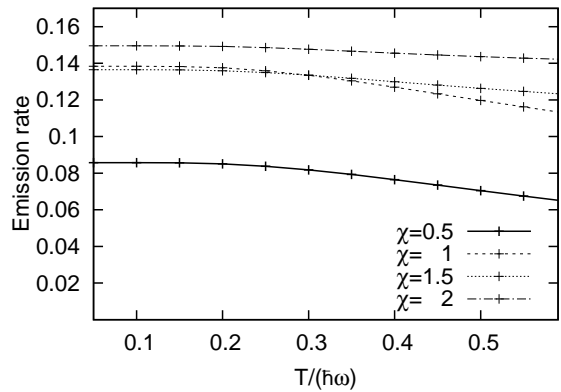}
\caption{ $3$rd harmonic emission rate in bilayer graphene $J_{3}$ (in
arbitrary units) as a function of the temperature scaled to photon energy
for various intensities. The wave frequency is $\protect\omega =$\ $0.05$\ $%
\mathrm{eV/}\hbar $, band gap energy $U=0.1$\textrm{\ }$\mathrm{eV}$. }
\end{figure}

For the clarification of the harmonics generation due to multiphoton
resonant excitation and particle-hole annihilation, from the coherent
superposition states at $\gamma \simeq 1$ initially we examine the emission
rate of the third harmonic. The fixed pump wave frequency is taken from
terahertz domain in Figs. 3-7. The emission rate dependence versus pump wave
strength defined by fixed value of parameter $\chi =1$ is demonstrated in
Fig. 3 for various gap energies. As is seen from this figure, for the field
intensities $\chi \succsim 1$ at the gap energy $U\approx 0.1$ $\mathrm{eV}$
we have a strong deviation from power law for the emission rate of the third
harmonic (in accordance with the perturbation theory $\sim \chi ^{3}$). In
Fig. 4 the dependence of emission rate on the energy gap is shown versus
pump wave intensity and gap energy at room temperature. In Fig. 5 third
harmonic emission rate is assumed as a function of the energy gap at various
intensities defined by $\chi $ at the same wave frequency. As a result, we
find the optimal parameters when the third harmonic emission rate is
significant: $U\approx 0.1$ $\mathrm{eV}$ and at the larger intensity for
the considered wave frequency.

To show the temperature dependence of the harmonic generation process at the
resonant excitation, in Figs. 6, 7 the third harmonic coherent emission rate
in bilayer graphene for various gap energies and various wave intensities
are plotted at the fixed wave frequency $\omega =$\ $0.05$\ $\mathrm{eV/}%
\hbar $. The Fig. 6 shows that at the fixed pump wave intensity for a case
of the found optimal value $U\approx 0.1$ $\mathrm{eV,}$ the third harmonic
emission rate is suppressed with increasing of the temperature. And, as is
seen from Fig. 7, at the same value $U\approx 0.1$ $\mathrm{eV}$, with the
temperature increase the emission rate of the third harmonic is strictly
diminished, as in intrinsic bilayer graphene at $U=0$ \cite{27} for a small
intensity only. The similar calculations for the intense pump wave or large
gap energy $U$ ($U\gg T$)\ have shown that emission rate exhibits a tenuous
dependence on the temperature.

So, in accordance to the results of Figs. 3-7, an intense radiation of the
third harmonic at the pump-wave-induced particle or hole acceleration and
annihilation in gapped graphene can be obtained at room temperatures with
the pump wave frequency in terahertz domain.

\section{Conclusion}

We find out the real/optimal values of energy gap produced by a constant
electric field, a wave intensity, and graphene temperature for the third
harmonic coherent radiation, using the microscopic quantum theory of
nonlinear interaction of a strong coherent radiation with a gapped bilayer
graphene. The band gap in this system is produced by a constant electric
field applied perpendicular to the surface of bilayer graphene. The closed
set of differential equations for the single-particle density matrix was
solved numerically for a gapped bilayer graphene in the vicinity of the $%
\zeta K$ points in the Brillouin zone, in nonadiabatic regime of interaction
when the Keldysh parameter is of the order of unity. In particular, the
obtained results for the third order harmonic generation show that optimal
values of characteristic parameters of the considering system make a gapped
bilayer graphene an effective medium for third--order nonlinear optical
effects at room temperatures for the pump wave frequencies in the terahertz
and far infrared domains.

\section{Appendix}

Here we present the set of differential equations for the quantities $N_{c}(%
\mathbf{p},t)$, $N_{\mathrm{v}}(\mathbf{p},t)$, $P(\mathbf{p},t)$ and final
relations for the transition dipole moment components and velocity operator.
Thus, we need to solve the closed set of differential equations for the
following quantities:%
\begin{equation*}
i\hbar \frac{\partial N_{c}(\mathbf{p},t)}{\partial t}-i\hbar e\mathbf{E}%
\left( t\right) \frac{\partial N_{c}(\mathbf{p},t)}{\partial \mathbf{p}}=
\end{equation*}%
\begin{equation}
\mathbf{E}\left( t\right) \mathbf{D}_{\mathrm{t}}\left( \mathbf{p}\right)
P^{\ast }(\mathbf{p},t)-\mathbf{E}\left( t\right) \mathbf{D}_{\mathrm{t}%
}^{\ast }\left( \mathbf{p}\right) P(\mathbf{p},t),  \label{20}
\end{equation}%
\begin{equation*}
i\hbar \frac{\partial N_{\mathrm{v}}(\mathbf{p},t)}{\partial t}-i\hbar e%
\mathbf{E}\left( t\right) \frac{\partial N_{\mathrm{v}}(\mathbf{p},t)}{%
\partial \mathbf{p}}=
\end{equation*}%
\begin{equation}
-\mathbf{E}\left( t\right) \mathbf{D}_{\mathrm{t}}\left( \mathbf{p}\right)
P^{\ast }(\mathbf{p},t)+\mathbf{E}\left( t\right) \mathbf{D}_{\mathrm{t}%
}^{\ast }\left( \mathbf{p}\right) P(\mathbf{p},t),  \label{21}
\end{equation}%
\begin{equation*}
i\hbar \frac{\partial P(\mathbf{p},t)}{\partial t}-i\hbar e\mathbf{E}\left(
t\right) \frac{\partial P(\mathbf{p},t)}{\partial \mathbf{p}}=
\end{equation*}%
\begin{equation*}
\left[ 2\mathcal{E}_{1}\left( \mathbf{p}\right) +\mathbf{E}\left( t\right) 
\mathbf{D}_{\mathrm{m}}\left( \mathbf{p}\right) -i\hbar \Gamma \right] P(%
\mathbf{p},t)
\end{equation*}%
\begin{equation}
+\mathbf{E}\left( t\right) \mathbf{D}_{\mathrm{t}}\left( \mathbf{p}\right) %
\left[ N_{\mathrm{v}}(\mathbf{p},t)-N_{c}(\mathbf{p},t)\right] ,  \label{21a}
\end{equation}

Taking into account Eq. (\ref{15}) with the spinor wave functions (\ref{4})
we obtain the transition dipole moment components in the ultimate forms:%
\begin{equation*}
D_{\mathrm{t}x}\left( \mathbf{p}\right) =-\frac{e\hbar }{2\mathcal{E}%
_{1}\left( \mathbf{p}\right) \sqrt{\mathcal{E}_{1}^{2}\left( \mathbf{p}%
\right) -\frac{U^{2}}{4}}}
\end{equation*}%
\begin{equation*}
\mathbf{\times }\left( \left[ \left( \frac{p^{2}}{2m}-m\mathrm{v}%
_{3}^{2}\right) \frac{\zeta p_{y}}{m}+\frac{\mathrm{v}_{3}}{m}p_{x}p_{y}%
\right] \right.
\end{equation*}%
\begin{equation}
\left. -i\frac{U}{2\mathcal{E}_{1}}\left\{ \left( \frac{p^{2}}{2m}+m\mathrm{v%
}_{3}^{2}\right) \frac{p_{x}}{m}-\frac{3\zeta \mathrm{v}_{3}}{2m}\left(
p_{x}^{2}-p_{y}^{2}\right) \right\} \right) ,  \label{27}
\end{equation}%
\begin{equation*}
D_{\mathrm{t}y}\left( \mathbf{p}\right) =-\frac{e\hbar }{2\mathcal{E}%
_{1}\left( \mathbf{p}\right) \sqrt{\mathcal{E}_{1}^{2}\left( \mathbf{p}%
\right) -\frac{U^{2}}{4}}}
\end{equation*}%
\begin{equation*}
\times \left( \left[ \left( -\frac{p^{2}}{2m}+m\mathrm{v}_{3}^{2}\right) 
\frac{\zeta p_{x}}{m}+\frac{\mathrm{v}_{3}}{2m}\left(
p_{x}^{2}-p_{y}^{2}\right) \right] \right.
\end{equation*}%
\begin{equation}
\left. -i\frac{U}{2\mathcal{E}_{1}}\left\{ \left( \frac{p^{2}}{2m}+m\mathrm{v%
}_{3}^{2}\right) \frac{p_{y}}{m}+\frac{3\zeta \mathrm{v}_{3}}{m}%
p_{x}p_{y}\right\} \right) .  \label{28}
\end{equation}%
Using Eq. (\ref{15}), the total mean dipole moments $D_{x,y\mathrm{m}}\left( 
\mathbf{p}\right) =D_{x,y\mathrm{m}}\left( 1,\mathbf{p}\right) -D_{x,y%
\mathrm{m}}\left( -1,\mathbf{p}\right) $ can be represented as:%
\begin{equation*}
D_{x\mathrm{m}}\left( \mathbf{p}\right) =-\frac{e\hbar U}{2\mathcal{E}%
_{1}\left( \mathbf{p}\right) \left( \mathcal{E}_{1}^{2}\left( \mathbf{p}%
\right) -\frac{U^{2}}{4}\right) }
\end{equation*}%
\begin{equation}
\mathbf{\times }\left[ \left( \frac{p^{2}}{2m}-m\mathrm{v}_{3}^{2}\right) 
\frac{\zeta p_{y}}{m}+\frac{\mathrm{v}_{3}}{m}p_{x}p_{y}\right] ,  \label{29}
\end{equation}%
\begin{equation*}
D_{y\mathrm{m}}\left( \mathbf{p}\right) =-\frac{e\hbar U}{2\mathcal{E}%
_{1}\left( \mathbf{p}\right) \left( \mathcal{E}_{1}^{2}\left( \mathbf{p}%
\right) -\frac{U^{2}}{4}\right) }
\end{equation*}%
\begin{equation}
\times \left[ \left( -\frac{p^{2}}{2m}+m\mathrm{v}_{3}^{2}\right) \frac{%
\zeta p_{x}}{m}+\frac{\mathrm{v}_{3}}{2m}\left( p_{x}^{2}-p_{y}^{2}\right) %
\right] .  \label{30}
\end{equation}%
The velocity operator is defined by the relation $\widehat{\mathbf{v}}%
_{\zeta }=\partial \widehat{H}/\partial \widehat{\mathbf{p}}$. After the
simple calculations for the effective $2\times 2$ Hamiltonian (\ref{1}), the
velocity operator in components can be presented by the expressions:%
\begin{equation}
\widehat{\mathrm{v}}_{\zeta x}=\zeta \left( 
\begin{array}{cc}
0 & -\frac{1}{m}\left( \zeta \widehat{p}_{x}-i\widehat{p}_{y}\right) +%
\mathrm{v}_{3} \\ 
-\frac{1}{m}\left( \zeta \widehat{p}_{x}+i\widehat{p}_{y}\right) +\mathrm{v}%
_{3} & 0%
\end{array}%
\right) ,  \label{51}
\end{equation}%
\begin{equation}
\widehat{\mathrm{v}}_{\zeta y}=i\left( 
\begin{array}{cc}
0 & \frac{1}{m}\left( \zeta \widehat{p}_{x}-i\widehat{p}_{y}\right) +\mathrm{%
v}_{3} \\ 
-\frac{1}{m}\left( \zeta \widehat{p}_{x}+i\widehat{p}_{y}\right) -\mathrm{v}%
_{3} & 0%
\end{array}%
\right) .  \label{52}
\end{equation}

\begin{acknowledgments}
The authors are deeply grateful to prof. H. K. Avetissian for permanent discussions and valuable recommendations.
This work was supported by the RA MES Science Committee.
\end{acknowledgments}

\end{document}